\documentclass[11pt,a4paper]{article} 
\usepackage{jheppub}

\usepackage{fancyhdr}
\usepackage{amssymb,amsmath}
\usepackage{tikz}
\usepackage{graphicx}
\usepackage{slashed}
\usepackage{soul}

\definecolor{yellow text}{rgb}{0.8,0.8,0}

\newcommand{\eq}[1]{Eq.~(\ref{#1})}
\newcommand{\eqs}[2]{Eqs.~(\ref{#1}) and (\ref{#2})}
\newcommand{\eqss}[3]{Eqs.~(\ref{#1}), (\ref{#2}) and (\ref{#3})}

\newcommand{\eV}{\mathinner{\mathrm{eV}}}
\newcommand{\keV}{\mathinner{\mathrm{keV}}}
\newcommand{\MeV}{\mathinner{\mathrm{MeV}}}
\newcommand{\GeV}{\mathinner{\mathrm{GeV}}}
\newcommand{\TeV}{\mathinner{\mathrm{TeV}}}

\def\l{\left}
\def\r{\right}
\def\pl{{\rm P}}

\def\bea{\begin{eqnarray}}
\def\eea{\end{eqnarray}}
\def\beq{\begin{equation}}
\def\eeq{\end{equation}}

\title{Singlet Portal Extensions of the Standard Seesaw Models to a Dark Sector with
Local Dark Symmetry}

\author[a]{Seungwon Baek}
\author[a]{P. Ko} 
\author[a]{Wan-Il Park}

\affiliation[a]{School of Physics, KIAS, \\ Seoul 130-722, Korea}

\emailAdd{sbaek1560@gmail.com}
\emailAdd{pko@kias.re.kr}
\emailAdd{wipark@kias.re.kr}

\arxivnumber{}

\abstract{
Assuming ({\it i}) the seesaw mechanism works for neutrino masses and mixings, 
({\it ii}) dark matter is absolutely stable due to unbroken $U(1)_X$ dark gauge symmetry, 
and ({\it iii}) the singlet fields ($H^\dagger H$ and $N_R$) are portal to the dark sector, 
we construct a simple model which  is consistent with all the cosmological observations 
as well as terrestrial experiments available as of now, including  leptogenesis, 
extra dark radiation of $\sim 8 \%$ (resulting in $N_{\rm eff} = 3.130$ the effective number of neutrino species), 
Higgs inflation, small and large scale structure formation, and current relic density of 
scalar dark matter ($X$).  The electroweak vacuum of this model is stable up to 
Planck scale for $m_H = 125$ GeV without any other new physics.
The Higgs signal strength is equal to one as in the standard model for unbroken $U(1)_X$ case 
with a scalar dark matter,  but it could be less than one independent of decay channels if the dark matter is a dark sector 
fermion or if $U(1)_X$ is spontaneously broken, because of a mixing with 
a new neutral scalar boson in the models.   Detailed study of Higgs properties at the 
LHC would shed light on the models described in this work. 
}
\begin{document}
\maketitle
\section{Introduction}

The standard model (SM) has passed many tests from various experiments from atomic physics 
scale up to a couple of TeV energy range. Still it is well known that the SM
has to be extended in order to accommodate the neutrino masses and mixings,
baryon number asymmetry (BAU) and cold dark matter (CDM) of the universe.
The most economic explanation for the first two problem would be leptogenesis 
\cite{leptogenesis,lepto_review},  
whereas there are many models for CDM in particle physics \cite{dm_review,Bergstrom:2000pn,Bertone:2004pz,Feng:2010gw}. 

For CDM physics, one of the puzzles is how CDM can be absolutely stable or 
very long lived. If unstable, the lifetime of CDM should be far longer than the
age of the universe, say $\tau \gtrsim 10^{26-30}$ sec \cite{Ackermann:2012qk}. 
Otherwise its decay would produce too much $X$($\gamma$)-ray or neutrino flux 
to match observation.   Still this lower bound of the CDM lifetime is far less than
the lower bound on the proton lifetime, the reason of which still remains
one of the mysteries in particle physics. 

The required longevity of the dark matter (DM) can be guaranteed by a symmetry.
If the symmetry is global, it can be broken by gravitational effects, and there can 
be dangerous operators suppressed by Planck scale ($M_\pl$), such as
\beq
- \mathcal{L}_{\rm decay} = \left\{
\begin{array}{ll}
 \frac{\lambda_{X, \rm non}}{M_\pl} ~ X F_{\mu \nu} F^{\mu \nu} & 
 \textrm{for bosonic~DM} \ X
\\
&
\\
 \frac{\lambda_{\psi, \rm non}}{M_\pl} ~\overline{\psi} \l( \slashed{D} \ell_{Li} \r) H^\dag & 
 \textrm{for fermionic~DM} \ \psi
 \end{array}
 \right.
 \eeq
where $\lambda_{X, \rm non} = \mathcal{O}(e^2)$ and $\lambda_{\psi, \rm non} \sim \mathcal{O}(1)$ 
are the couplings associated with the non-renormalizable operators, $X$ and $\psi$ are bosonic 
and fermionic dark mater candidates,  and $\ell_{Li}$ and $H$ are the SM lepton 
and Higgs, respectively.
In this case, dark matter can not be stable enough unless the mass of DM is small enough, e.g.,  
$m_X \lesssim \mathcal{O}(10) \keV$ for bosonic CDM $X$, or 
$m_\psi \lesssim \mathcal{O}(1) \GeV$ for fermionic CDM $\psi$. 
It may be possible to have such a light dark matter though it may not be 
theoretically so natural. Axion or keV scale sterile neutrinos are good examples 
of DM whose longevity is guaranteed by some global symmetries.   
On the other hand, the argument above implies that it is highly unlikely that an electroweak (EW) 
scale CDM is long lived or stable due to some global symmetries. 

Contrary to  global symmetry, local symmetry other than the SM gauge group 
often appears in theories beyond the SM, and would guarantee the absolute stability of dark 
matter if it is unbroken  \footnote{If a $Z_2$ discrete symmetry is the remnant of a broken 
local symmetry, it can be used to guarantee the stability of dark matter as often appears 
in literature. However the realistic model should  contain extra fields and couplings as 
discussed in Ref.~\cite{koz2,koz2-1,progress_broken}.}. 
For example, gauge groups in superstring 
theory have very large ranks, e.g. $SO(32)$ or $E_8 \times E_8^{'}$.  At low energy,
these gauge groups may be broken into a product of the SM gauge group 
($SU(3)_C \times SU(2)_L \times U(1)_Y$) and some other group, the latter of which
may be able to play a role of dark gauge group we consider in this paper.   
The presence of an unbroken extra local symmetry implies the existence of massless 
gauge boson(s). Since it is a carrier of long range force \footnote{We assume that the 
local dark symmetry is not confining. The confining nonAbelian hidden sector gauge 
interaction was considered in Ref.~\cite{ko_hidden_qcd1,ko_hidden_qcd2,ko_hidden_qcd_proceeding1,ko_hidden_qcd_proceeding2,ko_hidden_qcd_proceeding3,ko_hidden_qcd_proceeding4}.}, 
the massless dark gauge boson(s) could have significant effects on structure formation 
via self-interactions of  dark matter \cite{Carlson:1992fn,deLaix:1995vi,Spergel:1999mh}.
On one hand, the dark gauge interaction is highly constrained by various properties of small and large scale dark matter halos \cite{Buckley:2009in}.
On the other hand, it can provide a solution to small scale puzzles of the collisionless CDM scenario (e.g. cored density profiles \cite{de Blok:1997ut,Oh:2010ea,deNaray:2011hy,Walker:2011zu} and low concentrations of massive sub-halos \cite{BoylanKolchin:2011dk}) without conflicting with constraints from large scale structure \cite{Vogelsberger:2012ku}.
The massless dark gauge boson(s) could also contribute to the radiation density of the Universe in addition to thermal relic neutrinos of 3 species.
Recent WMAP 9-year data analysis showed that the number of relativistic degrees of freedom is \cite{Hinshaw:2012fq}  
\beq
N_{\rm eff}^{\rm obs} = 3.84 \pm 0.40 \ {\rm at} \ 68 \% \ {\rm CL}.
\eeq
Although it is consistent with the case of three active standard model neutrinos only ($N_{\rm eff}^{\rm SM} = 3.046$), some amount of extra radiation is still allowed and it could be from either light sterile neutrino \cite{Abazajian:2012ys} or hidden photon 
\cite{Holdom:1985ag} or axion \cite{Peccei:1977hh,Peccei:1977ur,Weinberg:1977ma,Wilczek:1977pj,Kim:1979if,Shifman:1979if,Zhitnitsky:1980tq,Dine:1981rt}. There are considerable amount of literatures on these possibilities. 

Meanwhile, dark sector can communicate with the SM sector via Higgs portal 
interactions ($H^\dagger H$) which are quite often used in the dark matter physics~\cite{EFT,EFT2,EFT_DLMQ,EFT_pseudo,Kim:2008pp,SFDM1} (see also \cite{Chu:2011be} where DM produced from SM particles via kinetic mixing and Higgs portals was analyzed).
Another possible portal interaction can be provided by heavy RH neutrinos
\footnote{
The operators $\tilde{H} l_{Li}$'s are also the SM gauge singlets as $H^\dagger H$, 
and could be a portal to another singlets from the hidden sector. 
We do not consider this because this operator is dim-$5/2$, and thus cannot have renormalizable couplings with composite operators made of the hidden sector fields 
charged under symmetry in the sector.  Instead, we trade $\tilde{H} l_{Li}$ with the lower dim
operators $N_{Ri}$'s in this paper.} which are 
singlet under  the SM gauge group
~\cite{Cosme:2005sb,Gu:2009yy,Gu:2009hj,An:2009vq,Chun:2010hz,Falkowski:2011xh}.  These singlet portal interactions are natural 
extensions of the SM in the framework of  renormalizable quantum field theory,  
and allow rich phenomenology in both dark matter and Higgs sector as we will show 
in the subsequent sections.

Based on this line of arguments and observations, in this paper we consider an extension 
of the SM  where a local $U(1)_X$ dark symmetry is introduced to guarantee the stability 
of dark matter.    The minimal particle contents and renormalizable interactions are 
completely fixed once portal interactions via Higgs and right-handed (RH) neutrinos are 
allowed.  These extensions allow a possibility to accommodate neutrino masses and 
mixings, leptogenesis for BAU, (a)symmetric dark matter and dark radiation.
In addition to these rich physics, Higgs inflation scenario can be also realized if large 
non-minimal couplings of scalar fields to gravity are introduced, and high enough 
reheating temperature after inflation sets a proper initial condition for the subsequent 
leptogenesis.

Before we proceed to the main discussions, let us make two comments. 
If we considered the spontaneously broken $U(1)_X$ case by introducing a new 
$U(1)_X$-charged scalar $\phi$ with $\langle \phi (x) \rangle \neq 0$,  
there would appear a new neutral scalar $h_X$ from the radial component of $\phi$.   
Then, this new neutral scalar will mix with the SM Higgs $h$, resulting into 2 neutral 
Higgs-like scalar bosons.  Since $h_X$ is the SM singlet scalar, the scalar boson sector 
will be similar to the case of Ref.~\cite{SFDM1}.   There, it was argued that the Higgs 
signal strength is always smaller than unity independent of  the decay channels of the 
Higgs boson, due to the mixing between the SM Higgs and 
the singlet scalar,  and also possible decays of scalar bosons into a pair of CDM's. 
This case would be  strongly disfavored if the current situation of enhanced 
$H\rightarrow \gamma\gamma$  remains there in the future analysis.  
Another issue  in the spontaneously broken dark symmetry case  is the stability or 
longevity of the dark matter candidate.  Nonrenormalizable  operators suppressed 
by some powers of (at least)   $M_\pl$ and even renormalizable operators for the 
scalar  dark matter case  would make the CDM decay in general, as long as electric 
charge, energy-momentum  and the total angular momentum are  conserved~\cite{koz2,progress_broken}.   One has to make a judicious choice of dark charge 
assignments in order to avoid these  problems. 
We postpone the detailed study of the spontaneously broken dark symmetry case
to the future \cite{progress_broken}, although we describe the qualitative features 
in Table~2 of Sec.~8. 

This paper is organized as follows. In Sec.~2, we define the model Lagrangian 
assuming the local gauge symmetry 
$SU(3)_C \times SU(2)_L \times U(1)_Y \times U(1)_X$  as the underlying gauge 
symmetry, where $U(1)_X$ is the unbroken dark symmetry which guarantees 
stability of the dark matter.  The right-handed neutrino singlet fields $N_{Ri}$'s are
also included for the seesaw mechanism and the leptogenesis. In Sec.~3, we 
consider various constraints on our model from large and small scale structure 
formation, vacuum stability and no Landau pole up to Planck scale, direct detection
cross section and indirect signatures after we identify the dark matter component 
in our model lagrangian. In Sec.~4, we calculate the amount of dark radiation within 
our model, which originates from massless dark photon. The leptogenesis 
from RH neutrino decays is discussed in Sec.~5. The possibility of Higgs inflation 
assisted with scalar dark matter is discussed in Sec.~6. Collider phenomenology of
Higgs boson and scalar dark matter is presented in Sec.~7. 
Some variations of our model are described in Sec.~8, with special emphasis on the 
nature of CDM and singlet portals, the number of Higgs-like neutral scalar bosons,  
extra dark radiation, and the Higgs signal strengths.  We discuss a few miscellaneous 
issues in Sec.~9, including the comparison of our model with other models  
in the literature and effects of nonrenormalizable operators. 
Finally we summarize the results in Sec.~10. Explicit expressions for thermally
averaged cross sections for the processes relevant to our discussions are presented
in Appendix.


\section{The Model}

As explained in Introduction, we assume that dark matter lives in a hidden sector, and it is stable 
due to unbroken local $U(1)_X$ dark gauge symmetry.  All the SM fields are taken to be $U(1)_X$ 
singlets. Assuming that the RH neutrinos are portals to the hidden sector, we need both a scalar ($X$) 
and a Dirac fermion ($\psi$) with the same nonzero dark charge (see Table~1). 
Then the composite operator $\psi X^\dagger$ becomes a gauge singlet and thus can 
couple to the  RH neutrinos $N_{Ri}$'s  
\footnote{If we did not assume that the RH neutrinos are portals to the dark sector, 
we did not have to introduce both $\psi$ and $X$ in the dark sector. This case is discussed 
in brief in Sec.~8.}.

With these assumptions, we can write the most general renormalizable Lagrangian as follows:   
\beq \label{Lagrangian}
\mathcal{L} = \mathcal{L}_{\rm SM} + \mathcal{L}_X + {\mathcal{L}_\psi} 
+ \mathcal{L}_{\rm kin-mix} +  \mathcal{L}_{\rm H-portal} + \mathcal{L}_{\rm RHN-portal}  
\eeq
where $\mathcal{L}_{\rm SM}$ is the standard model Lagrangian and  
\bea
\label{LX}
\mathcal{L}_X &=& {\l| \l( \partial_\mu + i g_X q_X \hat{B}'_\mu \r) X \r|^2} - \frac{1}{4} \hat{B}'_{\mu \nu} \hat{B}'^{ \mu \nu} - m_X^2 X^\dag X - \frac{1}{4} \lambda_X \l( X^\dag X \r)^2 ,
\\
\mathcal{L}_\psi &=& i \bar{\psi} \gamma^\mu \l( \partial_\mu + i g_X q_X \hat{B}'_\mu \r) \psi - m_\psi \bar{\psi} \psi ,
\\
\label{kin-mix}
\mathcal{L}_{\rm kin-mix} &=& - \frac{1}{2} \sin \epsilon \hat{B}'_{\mu \nu} \hat{B}^{\mu \nu} ,
\\
\label{Hportal}
\mathcal{L}_{\rm H-portal} &=& - \frac{1}{2} \lambda_{HX} X^\dag X H^\dag H ,
\\
\label{RHNportal}
- \mathcal{L}_{\rm RHN-portal} &=& \frac{1}{2} M_i \overline{N_{Ri}^C} N_{Ri} + \left[ Y_\nu^{ij} \overline{N_{Ri}} \ell_{Lj} H^\dag + \lambda^i \overline{N_{Ri}} \psi X^\dag + \textrm{H.c.} \right]. 
\eea
$g_X$, $q_X$, $\hat{B}'_\mu$ and $\hat{B}'_{\mu \nu}$ are the gauge coupling, $U(1)_X$ charge, the gauge field and 
the field strength tensor of the dark $U(1)_X$, respectively. 
$\hat{B}_{\mu \nu}$ is the gauge field strength of the SM $U(1)_Y$.
We assume 
\beq
m_X^2 > 0, \quad \lambda_X > 0, \quad \lambda_{HX} > 0
\eeq
so that the local $U(1)_X$ remains unbroken and the scalar potential is bounded from below 
at tree level \footnote{Quantum corrections to the scalar potential will be discussed in Sec.~3.2.}.

Either $X$ or $\psi$ is absolutely stable due to the unbroken local $U(1)_X$ gauge 
symmetry, and will be responsible for  the present relic density of nonbaryonic CDM.  
In our model, there is a massless dark photon which couples to the  SM $U(1)_Y$ gauge field 
by kinetic mixing.  One can diagonalize the kinetic terms by taking a linear transformation defined 
as \cite{Holdom:1985ag} \beq
\l(
\begin{array}{l}
\hat{B}_\mu \\ \hat{B}'_\mu
\end{array}
\r)
=
\l(
\begin{array}{cc}
1 / \cos \epsilon & 0
\\
- \tan \epsilon & 1
\end{array}
\r)
\l(
\begin{array}{l}
B_\mu \\ B'_\mu
\end{array}
\r).
\eeq 
In this basis, the SM $U(1)_Y$ gauge coupling is redefined as 
$g_Y = \hat{g}_Y / \cos \epsilon$, and hidden photon does not couple to the SM fields.
However, dark sector fields now couple to the SM photon and $Z$-boson.
In the small mixing limit, the couplings are approximated to 
\beq \label{DM-SM-int}
\mathcal{L}_{\rm DS-SM} = {{\bar{\psi} i \gamma^\mu \l[ \partial_\mu - i g_X q_X t_\epsilon \l( c_W A_\mu - s_W Z_\mu \r) \r] \psi}} + \l| \l[ \partial _\mu - i g_X q_X t_\epsilon \l( c_W A_\mu - s_W Z_\mu \r) \r] X \r|^2
\eeq
where $t_\epsilon = \tan \epsilon$, $c_W = \cos \theta_W$ and $s_W = \sin \theta_W$ with $\theta_W$ being the Weinberg angle.
Hence, dark sector fields charged under $U(1)_X$ can be regarded as mini-charged particles under electromagnetism after the kinetic mixing term is removed by a field redefinition,  Eq.~(2.8). 

Meanwhile, we can assign lepton number and $U(1)_X$ charge to RH neutrinos and dark fields 
as shown in Table~\ref{tab:charges}.
\begin{table}[htdp] 
\begin{center}
\begin{tabular}{|c||c|c|c|}
\hline
field & $N$ & $\psi$ & $X$ \\
\hline \hline
$q_L$ & 1 & 1 & 0 \\
\hline
$q_X$ & 0 & 1 & 1 \\
\hline
\end{tabular}
\end{center}
\caption{Lepton number and $U(1)_X$ charge assignment}
\label{tab:charges}
\end{table}
Then, the global lepton number is explicitly broken by Majorana mass terms for the RH neutrinos.
If $Y_\nu$ and $\lambda_i$ carry $CP$-violating phases, the decay of RH neutrinos can develop 
lepton number asymmetry in both of visible and dark sectors.  
Since $U(1)_X$ is unbroken, the asymmetry in the dark sector has a relation,
\beq \label{dpsiPlusdx}
Y_{\Delta \psi} + Y_{\Delta X} = 0
\eeq
where $Y_{\Delta i} \equiv (n_i - n_{\bar i})/s$ is the asymmetry between $i$ and $\bar{i}$ with 
$n_i$ and $s$ being the number density of $i$ and entropy density.

There are various physics issues involved in our model as listed below: 
\begin{itemize}
\item Small and large scale structure
\item Vacuum stability of Higgs potential
\item CDM relic density and direct/indirect DM searches
\item Dark radiation
\item Leptogenesis
\item Higgs inflation in case of a large non-minimal gravitational couplings
\end{itemize}
In other words, the model will be highly constrained, but astonishingly it turns out that our model can 
also explain various issues related to those physics in its highly constrained narrow parameter space 
without conflicting with any phenomenological, astrophysical and cosmological observations. 
It is highly nontrivial that our model can accommodate all these constraints in a certain parameter 
region, reminding us that our model was based on local gauge principle for the dark matter stability,
and assumption of singlet portals to the dark sector, by introducing only 3  new fields, $X, \psi$ and 
$\hat{B}'_\mu$. 

\section{Constraints}
Including the portal interactions, the presence of an unbroken local $U(1)_X$ in dark sector 
with kinetic mixing with the SM sector is subject to various phenomenological and cosmological
constraints.  In this section, we will take a look each of constraint or physics one by one.

\subsection{Structure formation}
\label{structure-form}
The presence of the dark matter self-interaction caused by nonzero charge of $U(1)_X$ could affect significantly the kinematics, shape and density  profile of dark matter halo, so it is constrained by, for example, the galactic dynamics \cite{Ackerman:2008gi}, ellipticity of dark matter halos \cite{MiraldaEscude:2000qt} and Bullet Cluster \cite{Randall:2007ph} (see also \cite{Ostriker:1999ee,Buckley:2009in,Loeb:2010gj,Vogelsberger:2012ku}).
For a velocity-dependent self-interaction, the transfer cross section of the dark matter self-interaction, defined as $\sigma_T = \int d \Omega \l( 1 - \cos \theta \r) \frac{d \sigma}{d \Omega}$, is upper-bounded as \cite{Vogelsberger:2012ku}
\beq \label{sigmaT-obs}
\l. \frac{\sigma_T^{\rm obs}}{m_{\rm dm}} \r|_{v=10 {\rm km/s}} \lesssim 35 \ {\rm cm^2/g}.
\eeq 
Interestingly, it was shown that, if $\sigma_T^{\rm obs}$ is close to the bound, it can solve the core/cusp problem \cite{Oh:2010ea} and ``too big to fail'' problem \cite{BoylanKolchin:2011dk} of the standard collisionless CDM scenario \cite{Vogelsberger:2012ku}.  

In our model, for both $\psi$ and $X$ the self-interaction cross section with a massless dark photon is given by \cite{Feng:2009mn} 
\beq
\sigma_T \simeq \frac{16 \pi \alpha_X^2}{m_{X(\psi)}^2 v^4} 
\ln \l[ \frac{m_{X(\psi)}^2 v^3}{(4 \pi \rho_X\alpha_X^3)^{1/2}} \r]
\eeq
where $v$ and $\rho_X$ are the velocity and density of the dark matter at the region of interest \footnote{There are other $t$-channel scatterings 
of $X$-$X^\dag$ (Higgs and $Z$-boson mediations) and the contact interaction-$\lambda_X$, but they don't 
have large enhancement caused by small velocity.}.
We take $v = 10 {\rm km/sec}$ and $\rho_X = 3 \GeV /{\rm cm}^3$.
Then, compared to \eq{sigmaT-obs}, dark interaction is constrained as
\beq \label{DMstructure-const}
\alpha_X \lesssim 5 \times 10^{-5} \l( \frac{m_{X(\psi)}}{300 \GeV} \r)^{3/2}
\eeq
where we approximated the log factor to $41$.
Either $X$ or $\psi$, which is lighter than the other, poses a stronger constraint on $\alpha_X$.
Note that $\psi$ couples only to dark photon at low energy, and the thermally-averaged  
annihilation cross section of 
$\psi$ is found to be 
\beq \label{sv-psi}
\langle \sigma v \rangle_{\rm ann}^\psi \approx \frac{\pi \alpha_X^2}{2 m_\psi^2} .
\eeq
The abundance of $\psi$ at freeze-out is 
\beq \label{mYpsi}
\left. \frac{m_\psi n_\psi}{s} \right|_{T_{{\rm f},\psi}} 
= 3.79 \l( \frac{g_*(T_{{\rm f}, \psi})^{1/2}}{g_{*S}} \r) \l( \frac{m_\psi}{T_{{\rm f},\psi}} \r) \frac{1}{\langle \sigma v \rangle_{\rm ann}^\psi M_\pl}
\simeq  \frac{\langle \sigma v \rangle_{\rm ann}^{\rm th}}{\langle \sigma v \rangle_{\rm ann}^\psi} \l( \frac{m_{\rm dm} n_{\rm dm}}{s} \r)_{\rm obs}
\eeq
where
\beq \label{sv-th}
\langle \sigma v \rangle_{\rm ann}^{\rm th} \simeq 6 \times 10^{-26} {\rm cm}^3 / {\rm sec}
\eeq
is the thermally-averaged annihilation cross section which gives the right amount of 
present dark matter relic density
\footnote{
Since a dark matter charged under an unbroken symmetry can annihilate only with its anti-particle which constitutes the half of the whole CDM relic density, $\langle \sigma v \rangle_{\rm ann}^{\rm th}$ is larger than the one for a charge-neutral dark matter by a factor 2.
} 
corresponding to 
\beq
\l( \frac{m_{\rm dm} n_{\rm dm}}{s} \r)_{\rm obs} \simeq 2 \times 10^{-10} \GeV.
\eeq
In the far right-hand side of \eq{mYpsi}, we used the fact that , even if $\langle \sigma v \rangle _ {\rm ann}^\psi$ varies by several orders of magnitude, $m_{\rm dm} / T_{\rm f}$ is changed only by a factor of $\mathcal{O}(1)$. 
The constraint, \eq{DMstructure-const}, implies that 
\beq \label{psi-ann}
\frac{\langle \sigma v \rangle_{\rm ann}^{\rm th}}{\langle \sigma v \rangle_{\rm ann}^\psi} \gtrsim 3.5 \times 10^4
\times \l\{
\begin{array}{lcc}
\l( \frac{1 \TeV}{m_X} \r)^3 \l( \frac{m_\psi}{1 \TeV} \r)^2 & {\rm for} & m_X < m_\psi
\\
\l( \frac{1 \TeV}{m_\psi} \r) & {\rm for} & m_\psi < m_X.
\end{array}
\r.
\eeq
Hence, if $\psi$ were stable, it would be over-abundant at present.

In order to avoid the over-closing by $\psi$, we assume
\beq \label{CDM-const}
m_\psi > m_X
\eeq
so that $\psi$ can decay through the virtual RH neutrinos.
The decay rate of $\psi$ is given by 
\beq
\Gamma_\psi \simeq \Gamma_{\psi \to \nu X} + \Gamma_{\psi \to \nu X h}
\eeq
where
\bea
\Gamma_{\psi \to \nu X} &\simeq& \frac{\lambda_1^2}{16 \pi} \frac{\tilde{m}_\nu}{M_1} m_\psi \l( 1 - \frac{m_X^2}{m_\psi^2} \r)^2 ,
\\
\Gamma_{\psi \to \nu X h} &\simeq& \frac{1}{48 \pi^2} \l( \frac{m_\psi^2}{v_H^2} \r) \Gamma_{\psi \to \nu X}
\eea
with $\tilde{m}_\nu \equiv Y_\nu^2 v_H^2 / M_1$ and $v_H=174 \GeV$ being respectively a contribution to the neutrino mass matrix and the vev of Higgs field. 
The present CDM relic density poses the strongest constraint on $\Gamma_\psi$, 
as we will see in a moment.

\eq{dpsiPlusdx} implies that, even if the asymmetry between $\psi$ and $\bar{\psi}$ may arise in the 
decay of RH neutrinos,  once $\psi$ decays, the dark matter composed of $X$ and $X^\dag$ becomes 
totally symmetric irrespective of its origin. 
If $\psi$ decays before the thermal component of $X$ freezes out, the $X$'s coming 
from the decay of $\psi$ thermalize, which makes the number density $n_X$ return to 
that of thermal equilibrium.  The present relic density in this case is determined by the 
thermal relic, hence the annihilation cross section should be the one in \eq{sv-th} 
(``symmetric thermal'' case).
On the other hand, if $\psi$ decays after the thermal freeze-out of $X$, the annihilation cross section should be larger than the one for thermal relics (i.e., \eq{sv-th}) so that the non-thermal freeze-out to provide a right amount of relic density (``symmetric non-thermal'' case).
In this case, the required background temperature when $\psi$ decays is determined by the annihilation cross section of $X$.

In our model, the pair annihilation of $X$-$X^\dag$ can be controlled by the Higgs portal interaction $\lambda_{HX}$ which leads to $s$-wave annihilations. 
It freezes out at a temperature $T_f \sim m_X / 20$.
However dark matter can still be in kinetic equilibrium with thermal background at a 
lower temperature due to scatterings to SM particles.   The scattering is mediated by 
photon and Higgs thanks to the kinetic mixing and Higgs portal interaction.
The transfer cross section of photon-mediation is such that 
$\sigma_T \propto \epsilon^2 / T^2$.
Although the associated scattering could be quite efficient at a low temperature, $\epsilon \ll 1$ make it less efficient. 
We found that, for $\epsilon \sim \mathcal{O}(10^{-9})$ which will be of our interest as described in section~3.3, the momentum transfer rate via photon is too small to keep kinetic equilibrium after freeze-out. 
In case of Higgs mediation, the kinetic equilibrium can be maintained by the scattering mainly to charm quark to a temperature of the charm quark mass scale \cite{Hofmann:2001bi}.
At a lower temperature, the scattering rate is too small.

\begin{figure}[ht] 
\centering
\includegraphics[width=0.5\textwidth]{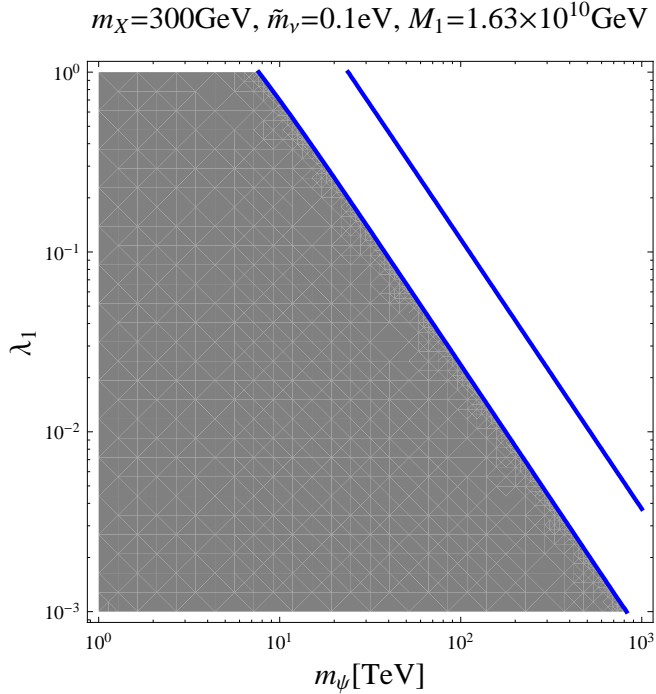}
\caption{
Parameter space for a right amount of dark matter relic density.
Contours correspond to the present dark matter relic density for 
$\langle \sigma v \rangle_{\rm ann}^X / \langle \sigma v \rangle_{\rm ann} ^{\rm th} = 1, 5$ (sold blue lines from right to left) with $m_X = 300 \GeV$, $\tilde{m}_\nu = 0.1 \eV$ and $M_1 = 1.63 \times 10^{10} \GeV$.
The gray region is excluded by XENON100 direct dark matter search as described in section~\ref{dd}.
}
\label{fig:non-th-fz}
\end{figure}

Hence, for $\lambda_{HX} \lesssim 1$ and electroweak scale $m_X$, the kinetic 
decoupling takes place  at a temperature $T_{\rm kd} \sim 1 \GeV$  
before QCD phase transition
\footnote{
As long as $X$ is decoupled before QCD-phase transition, the effect of the dark photon to the SM radiation at the time of BBN is negligible even though the dark photon is decoupled from $X$ at temperature $T \simeq 16 \MeV \l( \frac{5 \times 10^{-5}}{\alpha_X} \r) \l( \frac{m_X}{300 \GeV} \r)^{3/2}$ in our scenario \cite{Feng:2009mn}.
}.
If $\psi$ decays to $X$ abundantly, $X$ and $X^\dag$ would be able to re-annihilate even after freeze-out of the thermal annihilation until their number densities is reduced enough to stop the re-annihilation.
The abundance of $X$ and $X^\dag$ at the moment should be responsible for the present relic density 
of dark matter.   Hence, when $\psi$ decays at a temperature $T_{\rm d}$, the annihilation of $X$ should be frozen with a rate,
\beq
\Gamma_{\rm ann}(\tau_\psi) = n_X \langle \sigma v \rangle_{\rm ann,d}^X 
\eeq
with 
\beq
\l. \frac{n_X}{s} \right|_{T_{\rm d}}
\simeq 2 \times 10^{-12} \l( \frac{100 \GeV}{m_X} \r)
\eeq
being the present number density to entropy density ratio of $X$ that matches observation. 
$\langle \sigma v \rangle_{\rm ann,d}^X$ is the annihilation cross section of $X$ when $\psi$ decays.
Equating the annihilation rate to the expansion rate when $\psi$ decay, we find the decay temperature 
of $\psi$ to give a right amount of non-thermal relic density,
\beq
T_{\rm d} \equiv \l( \frac{\pi^2}{90} g_*(T_{\rm d}) \r)^{-1/4} \sqrt{\Gamma_\psi M_\pl} \simeq 9.2 \GeV \l( \frac{m_X}{300 \GeV} \r) \l( \frac{\langle \sigma v \rangle_{\rm ann}^{\rm th}}{\langle \sigma v \rangle_{\rm ann, d}^X} \r)
\eeq
where we used $g_*(T_{\rm d})=g_{*S}(T_{\rm d})=100$ and $M_\pl = 2.4 \times 10^{18} \GeV$ in the right-hand side of above equation.
This implies that 
\bea \label{lambda1-CDM-const}
\lambda_1^2 
&\simeq& 58.5 \l( \frac{0.1 \eV}{\tilde{m}_\nu} \r) \l( \frac{M_1}{10^9 \GeV} \r) \l( \frac{1 \TeV}{m_\psi} \r) \l( 1 - \frac{m_X^2}{m_\psi^2} \r)^{-2} \l[ 1 + \frac{1}{48 \pi^2} \l( \frac{m_\psi}{v_H} \r)^2 \r]^{-1} 
\nonumber \\
&& \phantom{58.5} \l( \frac{m_X}{300 \GeV} \r)^2 \l( \frac{\langle \sigma v \rangle_{\rm ann}^{\rm th}}{\langle \sigma v \rangle_{\rm ann,d}^X} \r)^2 .
\eea
Note that, if $\langle \sigma v \rangle_{\rm ann,d}^X = \langle \sigma v \rangle_{\rm ann}^{\rm th}$, $T_{\rm d}$ equal to or larger than $T_{\rm f}$ and corresponding $\lambda_1$ are fine. 
Note also that $T_{\rm d} > T_{\rm kd}$ unless $\langle \sigma v \rangle_{\rm ann,d}^X$ is larger than$\langle \sigma v \rangle_{\rm ann}^{\rm th}$ by at least two orders of magnitude.
However, as described in section~\ref{dd}, only $\langle \sigma v \rangle_{\rm ann}^X / \langle \sigma v \rangle_{\rm ann} ^{\rm th} \lesssim 5$ is allowed, so we can take $\langle \sigma v \rangle_{\rm ann,d}^X = \langle \sigma v \rangle_{\rm ann}^X$.
Fig.~\ref{fig:non-th-fz} shows contours for a right amount of relic density as a function of $\lambda_1$ and $m_\psi$.
In the figure, solid blue lines from right to left are for $\langle \sigma v \rangle_{\rm ann}^X / \langle \sigma v \rangle_{\rm ann} ^{\rm th} = 1$ and $5$ with $\tilde{m}_\nu = 0.1 \eV$, $M_1 =1.63 \times 10^{10} \GeV$ and $m_X = 300 \GeV$.

Shortly speaking, the existence of the massless dark photon constrains our model parameters to satisfy \eqs{DMstructure-const}{lambda1-CDM-const}.
They are from small/large scale structure formation and present dark matter relic density, respectively.

\subsection{Vacuum stability}
\label{vac-stab}
In the standard model, Higgs potential becomes unstable at an intermediate scale because of 
top loop contributions to the Higgs quartic couplings, though it depends on some of the standard 
model parameters, for example top pole mass and strong interaction \cite{Alekhin:2012py}.
Such instability can be cured if Higgs field couples to other scalar field(s) \cite{Lebedev:2012zw,EliasMiro:2012ay,Baek:2012uj}.  Depending on the existence of mixing between Higgs and additional scalar(s), 
tree-level and/or loop effects should be able to remove the instability.  
In our model, $X$ does not develop non-zero VEV, and the SM Higgs is not mixed with $X$. In this case, 
the loop-effect should be large enough to remove the vacuum instability of the SM Higgs potential.
Note that the Dirac neutrino mass terms also contribute to the RG-running of the Higgs quartic coupling.
However it is a negative contribution reflecting the fermionic nature of the right-handed neutrinos \cite{Rodejohann:2012px}.
Hence, in order not to make worse the vacuum instability up to Planck scale, we take 
\beq
Y_{\nu}^{ij} \lesssim 0.1,
\eeq
and ignore its contribution to the RG equation of Higgs quartic coupling.
Then, the relevant one-loop RG equations are
\beq
\beta_{\lambda_i} \equiv \frac{d \lambda_i}{d \ln \mu}
\eeq
where $i = H, HX, X$ and
\bea \label{lambdat-1loop-beta}
\label{lambdaH-1loop-beta}
\beta_{\lambda_H}&=& \frac{1}{16 \pi^2} \l[ 24 \lambda_H^2 + 12 \lambda_H \lambda_t^2 - 6 \lambda_t^4 - 3 \lambda_H \l( 3 g_2^2 + g_1^2 \r) + \frac{3}{8} \l( 2 g_2^4 + \l( g_2^2 + g_1^2 \r)^2 \r) + \frac{1}{8} \lambda_{HX}^2 \r] ,~~~~~~~
\\
\label{lambdaHS-1loop-beta}
\beta_{\lambda_{HX}} &=& \frac{\lambda_{HX}}{16 \pi^2} \l[ 2 \l( 6 \lambda_H + 3 \lambda_X + \lambda_{HX} \r) - \l( \frac{3}{2} \lambda_H \l( 3 g_2^2 + g_1^2 \r) - 6 \lambda_t^2 \r) \r] ,
\\
\label{lambdaS-1loop-beta}
\beta_{\lambda_X} &=& \frac{1}{16 \pi^2} \l[ \frac{1}{2} \lambda_{HX}^2 + 18 \lambda_X^2 \r]
\eea
in addition to the ones for the other SM couplings.
We solved 2-loop RGEs for SM couplings and 1-loop RGEs for non-SM couplings numerically, and found that the vacuum stability of Higgs potential and perturbativity of the couplings require 
\beq
0.2 \lesssim \lambda_{HX} \lesssim 0.6, \quad \lambda_X \lesssim 0.2.
\eeq

\subsection{Direct detection}
\label{dd}
In our model, dark matter couples to the SM particles via neutral SM gauge bosons (see \eq{DM-SM-int}) and Higgs portal, hence both type of interactions provide channels for dark matter direct searches.
In the case of gauge boson exchange, the spin-independent (SI) dark matter-nucleon scattering cross section via photon exchange provides a strong constraint on the kinetic mixing.
As can be seen from \eq{DM-SM-int}, our dark matter has a mini-electric charge, 
\beq
\epsilon_e = - \frac{g_X}{e} q_X c_W \tan \epsilon.
\eeq
For a scattering to a target atom with atomic number $Z$, the differential cross section of the Rutherford scattering of our dark matter  is given by
\beq \label{dsigma-dOmega}
\frac{d \sigma_A}{d \Omega} = \frac{\epsilon_e^2 \alpha_{\rm em}^2 Z^2 \mu_A^2}{4 m_X^4 v_{\rm cm}^4 \sin^4 (\theta_{\rm cm}/2)} F_A^2(q r_A)
\eeq
where $\mu_A \equiv m_X m_A / \l(m_X + m_A \r)$ with $m_A$ being the mass of the atom is the reduced mass, $v_{\rm cm}$ is the dark matter velocity at the center mass frame, and $\mathcal{F}_A(q r_A)$ is the form factor of the target atom with $q$ and $r_A$ being respectively the momentum transfer and effective nuclear radius.
The CM-frame scattering angle, $\theta_{\rm cm}$, is related to the nuclear recoil energy of the atom, $E_{\rm r}$, as
\beq
E_{\rm r} = \frac{\mu_A^2}{m_A} v^2 \l( 1 - \cos \theta_{\rm cm} \r)
\eeq
where $v$ is the lab velocity.
So, \eq{dsigma-dOmega} is expressed as
\beq \label{dsdE-th}
\l. \frac{d \sigma_A}{d E_{\rm r}} \r|_{\rm th}
= \frac{2 \pi \epsilon_e^2 \alpha_{\rm em}^2 Z^2}{m_A E_{\rm r}^2 v^2} \mathcal{F}_A^2(E_r).
\eeq
Experimentally, for the SI dark matter-nucleus scattering, the differential cross section with respect to the nucleus recoil energy is parameterized as
\beq \label{dsdE-exp}
 \l. \frac{d \sigma_A}{d E_{\rm r}} \r|_{\rm exp} = \frac{2 m_A Z^2}{\mu_p^2 v^2} \l( \sigma_p^{\rm SI} \r)_{\rm exp} \mathcal{F}_A^2(E_{\rm r})
\eeq
where $\mu_p = m_X m_p / \l( m_X + m_p \r)$ is the reduced mass of dark matter-proton system, $\l( \sigma_p^{\rm SI} \r)_{\rm exp}$ is the dark matter-proton scattering cross section constrained by experiments.   Note that the velocity dependence of \eq{dsdE-th} is 
the same as that of \eq{dsdE-exp}, and $\mathcal{F}^2_A(E_{\rm r}) / E_{\rm r}^2$ is a monotonically decreasing function for the range of $E_{\rm r}$ relevant in various direct detection experiments \cite{Lewin:1995rx}.
Hence, the kinetic mixing is  bounded from above as 
\beq \label{t-epsilon-bnd}
t_\epsilon < \l[ \frac{1}{\pi q_X^2 c_W^2 \alpha_X \alpha_{\rm em}} \r]^{1/2} \l( \frac{m_A}{\mu_p} \r) E_{\rm r}^{\rm T} \l( \sigma_p^{\rm SI} \r)^{1/2}
\eeq
where $E_{\rm r}^{\rm T}$ is the threshold recoil energy of a target atom at a given experiment.

\begin{figure}[ht] 
\centering
\includegraphics[width=0.45\textwidth]{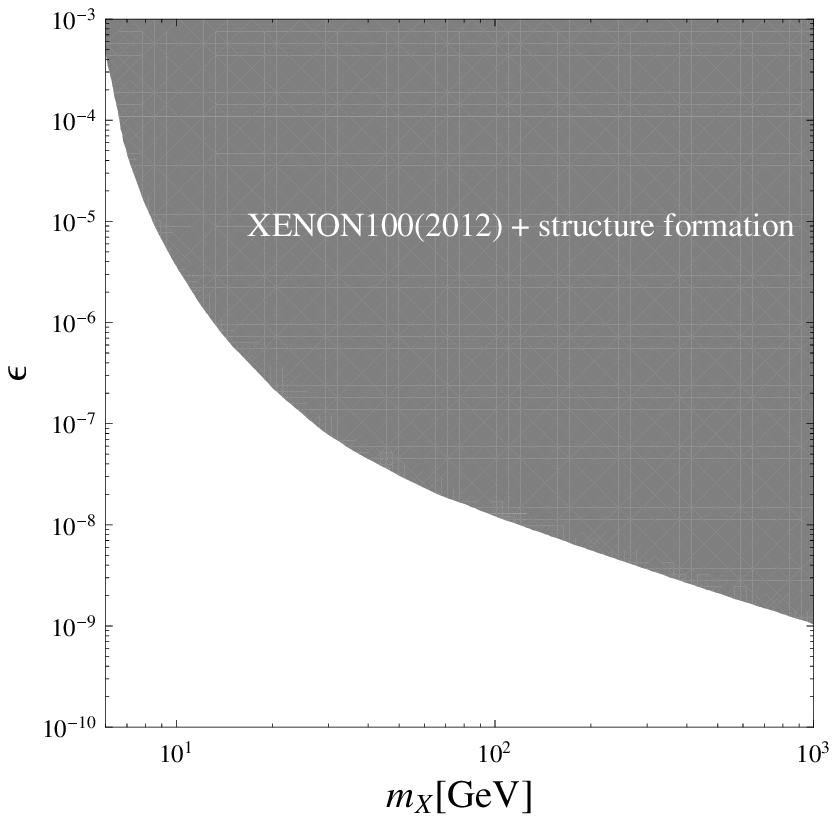}
\includegraphics[width=0.45\textwidth]{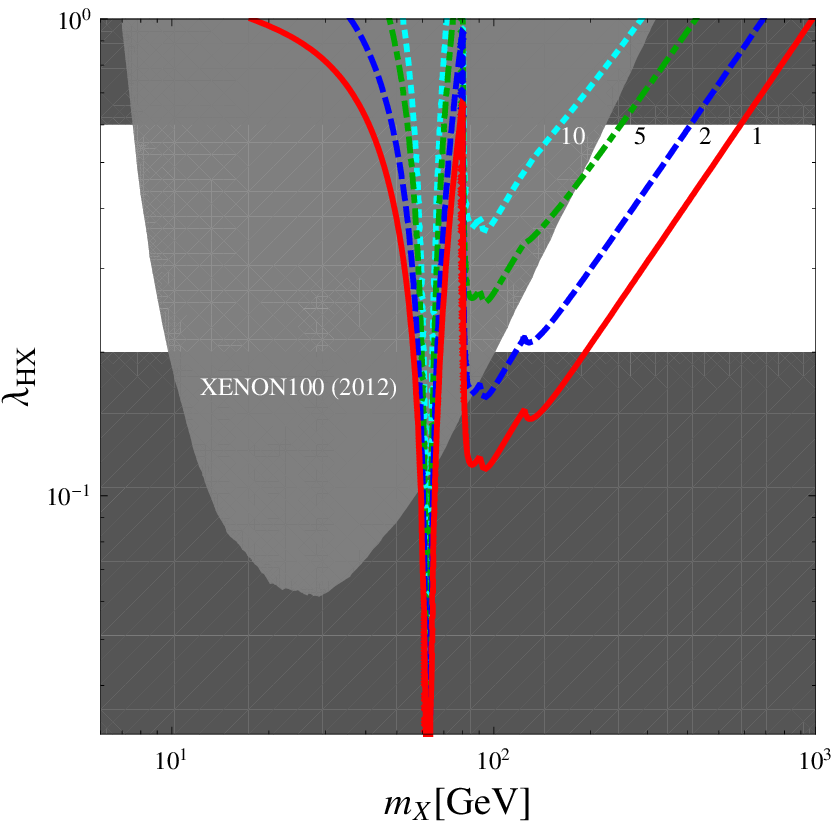}
\caption{
Left: XENON100(2012) bound on the kinetic mixing parameter $\epsilon$ as a function of $m_X$ for $\alpha_X$ 
given by the bound value of \eq{DMstructure-const}.
Right: XENON100 (2012) bound and contours of $\langle \sigma v \rangle_{\rm ann}^X 
/ \langle \sigma v \rangle_{\rm ann}^{\rm th}$ in ($\lambda_{HX}$, $m_X$) plane. 
$\langle \sigma v \rangle_{\rm ann}^{\rm th} \simeq 2 \times 10^{-36} {\rm cm}^2$ is the thermally 
averaged annihilation cross section giving the correct amount of dark matter relic density from thermal freeze-out. The gray region is excluded by the recent result from XENON100 \cite{Aprile:2012nq}. 
The lower and upper dark gray region is excluded by the vacuum stability of Higgs potential and perturbativity of couplings, respectively. 
The colored lines correspond to $\langle \sigma v \rangle_{\rm ann}^X / \langle \sigma v 
\rangle_{\rm ann}^{\rm th} = 1, 2, 5, 10$ from bottom to top.}
\label{fig:lHX-Xenon-bound}
\end{figure}

In the case of the Higgs portal interaction, the scattering cross section is 
\beq
\sigma_{\mathcal  {N}, H}^{\rm SI} = \frac{1}{\pi} m_{\rm r}^2 f_{\mathcal {N}, H}^2 
\eeq
where
\beq
f_{\mathcal {N}, H} = \frac{1}{8} \lambda_{HX} \frac{m_{\mathcal N}}{m_X m_H^2} f_{q,H}
\eeq
with 
\beq
f_{q,H} = \left[ \sum_{q=u,d,s} f_{Tq}^{\mathcal N} + \frac{2}{27} \sum_{q=t,b,c} f_{TG}^{\mathcal N} \right]
\eeq
and $f_{q}^{\mathcal N}$ and $f_{G}^{\mathcal N}$ being hadronic matrix elements with a scalar.
Based on the study on lattice \cite{Young:2009zb}, we take $f_q = 0.326$ here.
Currently, the strongest bound on $\sigma_p^{\rm SI}$ comes from XENON100 direct search experiment \cite{Aprile:2012nq} which has $E_{\rm r}^{\rm T} = 6.6 \keV$. 
Fig.~\ref{fig:lHX-Xenon-bound} shows how the kinetic mixing (left panel) and Higgs portal coupling (right panel) are limited by the experiment (gray region).
%
%
Also, depicted are thermally averaged annihilation cross sections (colored lines) and bounds 
from vacuum stability and perturbativity (dark gray regions).
In the left panel of the Fig.~2, we notice that, if small scale anomalies of structure formation are to be explained by the dark matter self-interaction, XENON100 direct search experiment constrains strongly the kinetic mixing as 
\beq
\epsilon \lesssim 10^{-9} - 10^{-4} \quad {\rm for} \quad 6 \GeV \lesssim m_X \lesssim 1 \TeV.
\eeq
From the right panel of Fig.~2, we also notice that direct search experiments already excluded $m_X \lesssim 80 \GeV$ except for the narrow resonance band around $m_X \simeq m_h/2$. 
In addition, for $m_X = \mathcal{O}(10^{2-3}) \GeV$, $\langle \sigma v \rangle_{\rm ann}^X$ can be larger than $\langle \sigma v \rangle_{\rm ann}^{\rm th}$ (the one for the right amount of thermal relic) by about an order of magnitude at most. 
For $m_h = 125 \GeV$, if top pole mass and strong coupling are respectively $m_t = 173.2 \GeV$ and $\alpha_s = 0.1184$, vacuum stability and perturbativity allows $m_X$ only in the range 
\beq \label{CDM-mass-window}
200 \GeV \lesssim m_X \lesssim 600 \GeV
\eeq
and annihilation cross section satisfying  
\beq
1 \leq \frac{\langle \sigma v \rangle_{\rm ann}^X}{\langle \sigma v \rangle_{\rm ann}^{\rm th}} \lesssim 5.
\eeq
This implies that the thermal relic can be reduced to abut $20$ \% of the present relic density at most, and asymmetrically produced non-thermal dark matter can saturate the present relic density. 
Note that the recent report on $E_\gamma \sim 130 \GeV$ line spectrum in Fermi-LAT $\gamma$-ray data is not achievable in our model since the branching fraction of the dark matter annihilation to photon(s) is of $\mathcal{O}(10^{-4}-10^{-3})$. 

\subsection{Indirect Signatures}
\label{sec:indirect-dec}
The dark interaction and kinetic mixing in our model should be highly suppressed as described in previous sections.
In addition, since $\alpha_X \lesssim 10^{-4}$ for $m_X \lesssim \mathcal{O}(1) \TeV$  (see \eq{DMstructure-const}), Sommerfeld enhancement factor, which is given by
\beq
S = \frac{\pi \alpha_X / v}{1 - e^{-\pi \alpha_X /v}},
\eeq
is $\mathcal{O}(1)$.
Hence it is difficult to expect detectable indirect signatures from the annihilation channels 
via dark interaction or kinetic mixing.

The possible indirect detection signatures comes from Higgs portal interactions,   
\[
X X^\dagger \rightarrow H^* \rightarrow f \bar{f}, V V,  \ \ {\rm or} \ \ X X^\dagger \rightarrow H H , 
\]
where $f$ and $V$ are the SM fermions and the weak gauge bosons, respectively.
These processes can produce a sizable continuum spectrum of photons, since the annihilation cross section can be larger than the value for thermal dark matter.
However, the recent data from Fermi LAT $\gamma$-ray search provides upper-bounds on various annihilation channels \cite{,Huang:2012yf}.
In our model, $W^+ W^-$ channel is dominant.
Taking into account the fact that an annihilation is possible only for $X$-$X^\dag$ pairs, the annihilation cross section is expected to be constrained at least as \cite{Huang:2012yf} 
\beq
\langle \sigma v \rangle_{XX^\dag \to W^+ W^-}^{\rm obs} \lesssim 2 \times 7.4 \times 10^{-26} {\rm cm}^3 / {\rm sec}
\eeq
for NFW dark matter profile.
Hence the total annihilation cross section is upper-bounded as 
\beq
\langle \sigma v \rangle_{\rm ann}^X \lesssim {\rm Br}(XX^\dag \to W^+ W^-)^{-1} \times 2 \times 7.4 \times 10^{-26} {\rm cm}^3 / {\rm sec}.
\eeq
In the allowed region of parameter space, that is, for $m_X = \mathcal{O}(10^{2-3}) \GeV$, we find ${\rm Br}(XX^\dag \to W^+ W^-) \sim 0.5$, and the allowed ratio of the annihilation cross section to the value for thermal relic is bounded as 
\beq
1 \leq \frac{\langle \sigma v \rangle_{\rm ann}^X}{\langle \sigma v \rangle_{\rm ann}^{\rm th}} \lesssim 5 .
\eeq
This constraint is similar to the one coming from the perturbativity bound shown in Fig.~\ref{fig:lHX-Xenon-bound}.

\section{Dark Radiation}  
\label{sec:dark-rad}
Dark photon can contribute to the radiation density of the present universe.
Its contribution is parameterized in terms of the extra relativistic neutrino species as
\beq
\Delta N_{\rm eff} = \frac{\rho_{\gamma'}}{\rho_\nu} = \frac{g_{\gamma'}}{(7/8) g_\nu} \l( \frac{T_{\gamma,0}}{T_{\nu,0}} \r)^4 \l( \frac{T_{\gamma',{\rm dec}}}{T_{\gamma,{\rm dec}}} \r)^4 \l( \frac{g_{*S}(T_{\gamma, 0})}{g_{*S}(T_{\gamma, \rm dec})} \r)^{4/3}
\eeq
where $\rho_{\gamma'}$ and $\rho_\nu$ are respectively the present energy densities of the dark photon and a neutrino species, $g_i$, $T_{i,0}$ and $T_{i, \rm dec}$ are respectively the degrees of freedom, the temperature at present and decoupling of the species, $i$, and $g_{*S}$ is the total SM degrees of freedom associated with entropy. 
Because of the energy injection to photons at the epoch of electron-positron pair annihilation which took place after neutrino decoupling, the photon is slightly hotter than neutrinos at present, resulting in the ratio of temperatures, $T_{\nu,0}/T_{\gamma,0} = \l( 4/11 \r)^{1/3}$.
In addition, dark matter is decoupled from the SM thermal bath at a temperature $T \sim 1 \GeV$ before QCD-phase transition while still in contact with dark photon.
Hence dark matter and dark photon are decoupled from the SM thermal bath simultaneously.
When it is decoupled, the temperature of dark photon is the same as that of photon.
Therefore, we find 
\beq
\Delta N_{\rm eff} =  \frac{g_{\gamma'}}{(7/8) g_\nu} 
\l( \frac{T_{\gamma,0}}{T_{\nu,0}} \r)^4 \l( \frac{g_{*S}(T_{\gamma, 0})}{
g_{*S}(T_{\gamma, \rm dec})} \r)^{4/3} \simeq 0.08
\eeq
where we used  $g_{\gamma'} = g_\nu = 2$, $g_{*S}(T_{\gamma, 0}) = 3.9$ and $g_{*S}(T_{\gamma, \rm dec})=75.75$.
The best fit value of observations is \cite{Hinshaw:2012fq}
\beq \label{Neff-obs}
N_{\rm eff}^{\rm obs} = 3.84 \pm 0.40 \ {\rm at} \ 68 \% \ {\rm CL}
\eeq
with SM expectation $N_{\rm eff}^{\rm SM} = 3.046$.
Therefore, in our model the contribution of dark photon to the radiation density 
at present is consistent with observation within about 2-$\sigma$ error, 
slightly improving the SM prediction in the right direction.

\section{Leptogenesis}
Our model allows production of lepton number asymmetries in both of visible and dark sectors via decays of heavy RH neutrinos.
If the mass of dark matter $X$ is much larger than proton mass and asymmetric generation of dark matter is responsible for the present relic density, the asymmetry of $\psi$ should be much smaller than that of lepton $\ell_i$.
However, the contribution to $X$ and $X^\dag$ from the decay of thermal symmetric component of $\psi$-$\bar{\psi}$ is dominant as described in section~\ref{structure-form}.  
The present relic density is then determined by thermal or non-thermal freeze-out of the annihilation of $X$-$X^\dag$, depending on the temperature when $\psi$ decays.
Considering asymmetric generation of dark matter in this circumstance is pointless.
However we still have to check if a right amount of lepton number asymmetry in the visible sector can be achieved.

The lepton number and $U(1)_X$ charges are assigned to relevant fields as shown in Table~\ref{tab:charges}.
Then, the global lepton number is explicitly broken by Majorana mass terms for the RH neutrinos.  
The lightest RH Majorana neutrino $N_1$ can decay into both the SM fields and the DM fields: 
\[
N_1 \rightarrow l_{Li} H^\dagger , \ \ \ \psi X^\dagger .
\]
With nonzero complex phases in $Y_\nu$ and $\lambda_i$ the decay can generate the $\Delta L$, $\Delta \psi$ and $\Delta X$ 
as
\footnote{For simplicity, we do not consider the case where the initial abundance of $N_1$ is negligible or zero.}
\beq \label{Yasym}
Y_{\Delta i} \equiv \frac{n_{\Delta_i}}{s} = \epsilon_i \eta_i Y_1^{\rm eq}(0)
\eeq
where $n_{\Delta_i}$ is the number density of a charge asymmetry associated with the field $i$, $s$ is the entropy density, $\epsilon_i$ and $\eta_i$ are asymmetry and wash-out effect of field $i$ from the decay of $N_1$, respectively, and $Y_1^{\rm eq}(0)=135 \zeta(3) / 4 \pi^4 g_*$ with $g_*(T \gg M_1) \sim 100$ being the number of relativistic degrees of freedom at 
a temperature well above the mass scale of the lightest RH neutrino ($M_1$).
For a hierarchical mass spectrum, $M_1 \ll M_{2,3}$, the asymmetries are given by \cite{Falkowski:2011xh}
\bea
\epsilon_L &\simeq& \frac{M_1}{8 \pi} \frac{{\rm Im} \l[ \l( 3 Y_\nu^* Y_\nu^T + \lambda^* \lambda \r) \mathbb{M}^{-1} Y_\nu Y_\nu^\dag \r]_{11}}{\l[ 2 Y_\nu Y_\nu^\dag + \lambda \lambda^* \r]_{11}} ,
\\
\epsilon_\psi &\simeq& \frac{M_1}{8 \pi} \frac{{\rm Im} \l[ \l( Y_\nu^* Y_\nu^T + \lambda^* \lambda \r) \mathbb{M}^{-1} \lambda \lambda^* \r]_{11}}{\l[ 2 Y_\nu Y_\nu^\dag + \lambda \lambda^* \r]_{11}}
\eea
where $\mathbb{M} = {\rm diag} \l(M_1, M_2, M_3 \r)$, and upper-bounded as \cite{Falkowski:2011xh,Davidson:2002qv}
\beq \label{gen-DI-bound}
\epsilon_L \leq \frac{3 M_1 m_\nu^{\rm max}}{16 \pi v_H^2} \times \l\{
\begin{array}{lcc}
1  & {\rm for} & {\rm Br}_L \gg {\rm Br}_\psi
\\
\sqrt{ \lambda_2^2 M_1 / \lambda_1^2 M_2} & {\rm for} & {\rm Br}_L \ll {\rm Br}_\psi
\end{array}
\r.
\eeq
with $m_\nu^{\rm max}$ being the mass of the heaviest left-handed neutrino.

The visible sector lepton number asymmetry, $\Delta L$, would end up the visible sector baryon number asymmetry via anomalous electroweak process \cite{Klinkhamer:1984di,Kuzmin:1985mm}. 
For simplicity, if we assume the visible sector lepton number asymmetry is dominated by a flavor, the late-time baryon number asymmetry is related to the lepton number asymmetry as \cite{lepto_review} 
\beq
Y_{\Delta B} = \frac{12}{37} Y_{\Delta L}.
\eeq
Then, the present observations of baryon number asymmetry 
can be matched if
\bea \label{YL}
Y_{\Delta L} &\simeq& 2.6 \times 10^{-10}.
\eea

The eventual outcome of leptogenesis via the decay of heavy RHN can be obtained by solving Boltzmann equations which involve effects of wash-out and transfer of the asymmetries between visible and dark sectors.
However, if the narrow-width approximation is valid, we can get much simpler picture.
The narrow-width approximation is valid if 
\beq
\frac{\Gamma_1^2}{M_1 H_1} \ll 1
\eeq
and $2 \to 2$ scattering between visible and dark sectors via heavy neutrino is ineffective, hence asymmetries in both sector evolves independently.
In this circumstance, the washout effect on asymmetry is mainly from the inverse decay.
If the washout effect is weak, i.e.,
\beq
{\rm Br}_i \frac{\Gamma_1}{H_1} \ll 1 ,
\eeq
the final asymmetry is directly related to the asymmetry from the decay of RHN.
Otherwise, there can be large reduction of the asymmetry.

The decay rate of RHN is 
\beq
\Gamma_1 = \frac{1}{16 \pi} \l( Y_{\nu 1}^2 + \lambda_1^2 \r) M_1 ,
\eeq
and the branching fractions to the SM and dark sectors are
\beq
{\rm Br}_L = \frac{Y_{\nu 1}^2}{Y_{\nu 1}^2 + \lambda_1^2}, \quad {\rm Br}_\psi = \frac{\lambda_1^2}{Y_{\nu 1}^2 + \lambda_1^2}.
\eeq
Hence
\beq \label{washout-cond}
{\rm Br}_i \frac{\Gamma_1}{H_1} = \frac{M_\pl}{16 \pi} \times \l\{
\begin{array}{lcc}
\tilde{m}_\nu / v_H^2 & {\rm for} & L
\\
\lambda_1^2 / M_1 & {\rm for} & \psi.
\end{array}
\r.
\eeq
For simplicity, we use narrow-width approximation from now on.
This implies 
\beq \label{narrow-wid-approx}
Y_{\nu_1}^2+ \lambda_1^2 \ll 16 \pi \l( \frac{M_1}{M_\pl} \r)^{1/2} \simeq 10^{-3} \l( \frac{M_1}{10^9 \GeV} \r)^{1/2}.
\eeq
Note that 
\beq \label{Ynu-mnu}
Y_{\nu_1}^2 = \frac{\tilde{m}_\nu M_1}{v_H^2} \simeq 3 \times 10^{-6} \l( \frac{\tilde{m}_\nu}{0.1 \eV} \r) \l( \frac{M_1}{10^9 \GeV} \r).
\eeq
So, $Y_{\nu 1}$ can not saturate the bound of \eq{narrow-wid-approx} for $M_1 \ll 10^{14} \GeV$ which we assumed in order not to worsen the vacuum instability of the SM Higgs potential.
Hence, for $\lambda_1$ saturating the bound, we always have ${\rm Br}_L \ll {\rm Br}_\psi$.
Combined with the constraint \eq{lambda1-CDM-const}, the narrow-width approximation can be achieved if  
\beq \label{mpsiLbnd}
m_\psi \gtrsim 94.3 \TeV  \l[ \l( \frac{0.1 \eV}{\tilde{m}_\nu} \r) \l( \frac{M_1}{10^9 \GeV} \r)^{1/2} \l( \frac{\langle \sigma v \rangle_{\rm ann}^{\rm th}}{\langle \sigma v \rangle_{\rm ann}^X} \r)^2 \r]^{1/3} \l( \frac{m_X}{300 \GeV} \r).
\eeq
Depending on the sizes of $Y_{\nu 1}$ and $\lambda_1$, there are various regimes of wash-out as analyzed in Ref.~\cite{Falkowski:2011xh}.
The purpose of this paper is not at the full analysis of leptogenesis, so here we simply show a working example in the following paragraph.

\begin{figure}[h] 
\centering
\includegraphics[width=0.5\textwidth]{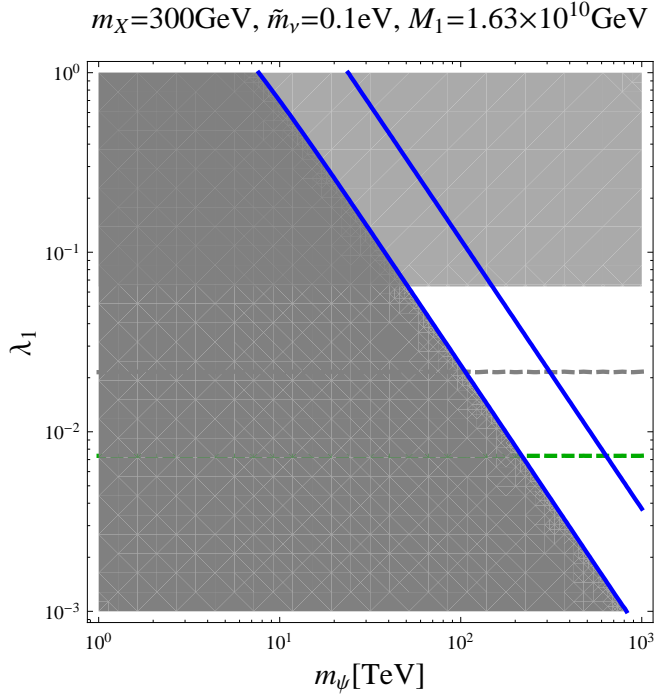}
\caption{Parameter space for right amounts of baryon number asymmetry and dark matter relic density at present. 
We used $m_X = 300 \GeV$, $\tilde{m}_\nu = 0.1 \eV$ and $\sqrt{Y_{\nu 2}^2 M_1 / Y_{\nu 1}^2 M_2} = 1$ corresponding to $M_1 = 1.63 \times 10^{10} \GeV$.
Dark gray region is excluded by XENON100 dark matter direct search experiment. 
In the light gray region, narrow-width approximation is not valid. 
The boarder of the light gray region and the gray dashed line correspond to 
$\lambda_1/ \sqrt{16 \pi \sqrt{M_1/M_\pl}} = 1, 1/3$, respectively.
Below the green line, $Y_{\nu 1} > \lambda_1$ for which our analysis is not valid.
The blue lines correspond to $\langle \sigma v \rangle_{\rm ann}^X / \langle \sigma v \rangle_{\rm ann}^{\rm th} = 1, 5$ from right to left.
}
\label{fig:para-space-narrow-wid-approx}
\end{figure}

If $\tilde{m}_\nu \sim 0.1 \eV$ and $\lambda_1 > Y_{\nu 1}$, both of visible and hidden sectors are in the strong washout regime.
The wash-out effects are given by \cite{Falkowski:2011xh,lepto_review} \beq \label{eta-strong}
\eta_L \simeq \frac{H_1}{\Gamma_1 {\rm Br}_L}, \quad \eta_\psi \simeq \frac{H_1}{\Gamma_1 {\rm Br}_\psi}
\eeq 
with the ratio between asymmetries,
\beq \label{dLdpsi}
\frac{Y_{\Delta L}}{Y_{\Delta \psi}} \simeq \frac{\epsilon_L {\rm Br}_L}{\epsilon_\psi {\rm Br}_\psi} \simeq \frac{\lambda_1 Y_{\nu 2}}{\lambda_2 Y_{\nu 1}}.
\eeq
Since $Y_{\Delta \psi}$ can be smaller or larger than $Y_{\Delta L}$ even though $m_X$ is much larger than proton mass, we can assume
\beq \label{YLeqYpsi}
\frac{\lambda_1 Y_{\nu 2}}{\lambda_2 Y_{\nu 1}} = 1.
\eeq
From \eqss{Yasym}{gen-DI-bound}{eta-strong}, the maximally expected late-time lepton number asymmetry is 
\beq \label{YLmax}
Y_{\Delta L}^{\rm max} 
= 1.6 \times 10^{-11} \l(\frac{M_1}{10^9 \GeV} \r) \l( \frac{\lambda_2^2 M_1}{\lambda_1^2 M_2} \r)^{1/2}.
\eeq
Hence the present baryon number asymmetry corresponding to \eq{YL} can be obtained if 
\beq \label{BAU-const}
\l(\frac{M_1}{10^9 \GeV} \r) \l( \frac{Y_{\nu 2}^2 M_1}{Y_{\nu 1}^2 M_2} \r)^{1/2} \simeq 16.3
\eeq
where we used \eq{YLeqYpsi} in the left-hand side of above equation.
Fig.~\ref{fig:para-space-narrow-wid-approx} shows a parameter space limited in our analysis.
In the figure, dark gray region is excluded by XENON100 direct dark matter search.
Narrow-width approximation is valid in the white region well below the light gray region (e.g., below the dashed gray line).
$\lambda_1 < Y_{\nu 1}$ below the green line.
Although a wider parameter space may be allowed, our analysis of leptogenesis in this section is limited only in the white region bounded by the dashed gray and green lines.
In the region, right amounts of baryon number asymmetry and dark matter relic density can be obtained as long as \eq{BAU-const} is satisfied.

So far, we have considered the lepton number asymmetry in the visible sector that comes from the decay of RH-neutrinos only.
However there is an additional contribution from the late-time decay of $\psi$ which also carries lepton number.
Since the decays of $\psi$ and $\bar{\psi}$ involve a virtual internal line of a Majorana RH-neutrino which decays eventually to a SM lepton and Higgs pair, both of decays produce equal amount of the same-sign lepton number asymmetry in the visible sector.
In addition, there is no dilution of the produced visible sector asymmetry due to inverse decay or transfer to the dark sector, since such processes are kinematically forbidden.
Hence, the contribution from those decays is 
\beq
\Delta (Y_{\Delta L}) = 2 \epsilon_L Y_\psi(T_{\rm fz}^\psi)
\eeq
where $T_{\rm fz}^\psi$ is the freeze-out temperature of the pair annihilation of $\psi$-$\bar{\psi}$ and we used $Y_\psi(T_{\rm d}) = Y_\psi(T_{\rm fz}^\psi)$ in the right-hand side of the above equation.  
The freeze-out abundance of $\psi$ is given by \cite{Kolb-Turner}
\beq
Y_\psi(T_{\rm fz}^\psi) 
= \frac{3.79 \l( \sqrt{8 \pi} \r)^{-1} g_*^{1/2} /g_{*S} x_{\rm fz}^\psi}{m_\psi M_\pl \langle \sigma v \rangle_{\rm ann}^\psi}
\simeq 0.05 \frac{x_{\rm fz}^\psi}{\alpha_X^2} \frac{m_\psi}{M_\pl }
\eeq
where $x_{\rm fz}^\psi \equiv m_\psi / T_{\rm fz}^\psi$ and we used $g_* = g_{*S} = 100$ and \eq{sv-psi} at the far right-hand side of the above equation.
Combining with \eqs{gen-DI-bound}{YL}, we find 
\beq
\frac{\Delta (Y_{\Delta L})}{Y_{\Delta L}} \simeq 2 \times 10^7 \frac{x_{\rm fz}^\psi}{\alpha_X^2} \frac{m_\psi}{M_\pl }
\frac{M_1 m_\nu^{\rm max}}{v_H^2} \times \l\{
\begin{array}{lcc}
1  & {\rm for} & {\rm Br}_L \gg {\rm Br}_\psi
\\
\sqrt{ \lambda_2^2 M_1 / \lambda_1^2 M_2} & {\rm for} & {\rm Br}_L \ll {\rm Br}_\psi.
\end{array}
\right.
\eeq 
%
As an example, we may take $\epsilon_L \sim 10^{-7}$.
Then, for  $\alpha_X = 10^{-5}$ and $m_\psi = 10^3 \TeV$, we find $x_{\rm fz}^\psi \simeq 2.2$ resulting in $\Delta (Y_{\Delta L}) / Y_{\Delta L} \simeq 0.3$. 
Therefore, depending on $\alpha_X$ and $m_\psi$, the decay of $\psi$ and $\bar{\psi}$ can be the origin of the baryon number asymmetry in the present universe even though the asymmetry between $\psi$ and $\bar{\psi}$ is absent. 

\section{Higgs Inflation}

In order for the leptogenesis described in the previous section to work, 
the temperature of the early universe should be high enough so that the lightest 
RHN can be in thermal equilibrium before it is decoupled.
This condition can be achieved if the reheating temperature of the primordial inflation
is high enough.  An intriguing possibility is so-called Higgs inflation 
\cite{Bezrukov:2007ep,Bezrukov:2010jz} which uses the SM Higgs as the inflaton 
equipped with a large non-minimal gravitational coupling.
As a variant, Higgs-scalar singlet system has been also considered in the literature \cite{Lerner:2009xg} (see also \cite{Clark:2009dc,Lebedev:2011aq,GarciaBellido:2011de}). 
Modulo the subtle issues of the unitarity problem \cite{Burgess:2009ea,Barbon:2009ya,Burgess:2010zq,Hertzberg:2010dc,Lerner:2011it}, our model 
indeed allows inflation along Higgs direction since Higgs potential is stabilized by the 
help of a coupling to the singlet scalar $X$.
The model parameters relevant to inflation are $\lambda_{HX}$, $\lambda_X$ and 
the Higgs quartic coupling in addition to the large non-minimal couplings (say $\xi_i$).
As free parameters, we can adjust $\xi_i$s for given set of quartic couplings while 
satisfying requirements on the inflationary observables under the assumption of the 
positivity of quartic couplings (see \cite{Lebedev:2012zw} for example).
Hence the physics involved in inflation does not pose any new constraint other than ones
described in previous sections if inflation takes place along Higgs direction, and 
the Higgs inflation along with a singlet scalar can be realized.

It turned out that the reheating temperature after Higgs inflation is around 
$\mathcal{O}(10^{13-14}) \GeV$ \cite{Bezrukov:2008ut}.
It is high enough to populate the lightest RHN in thermal bath.
Therefore, Higgs inflation sets the initial condition for the leptogenesis.

\section{Higgs and DM phenomenology at colliders}

The Higgs boson in our model could decay into a pair of scalar DM's through $\lambda_{HX}$ term if kinematically allowed.
However, as shown in Fig.~\ref{fig:lHX-Xenon-bound}, dark matter direct search allows only $m_X \sim m_h /2$ with $\lambda_{HX} \lesssim  10^{-1}$ even though SM Higgs may not suffer from vacuum instability problem.
If it is allowed, the decay rate of Higgs to dark matter is 
\beq
\Gamma_{h \to XX^\dag} = \frac{\lambda_{HX}^2}{128 \pi} \frac{v_H^2}{m_h} \l( 1 - \frac{4 m_X^2}{m_h^2} \r)^{1/2} ,
\eeq
and the signal strength ($\mu$) of SM Higgs searches at collider experiments is given by
\beq
\mu = 1 - \frac{\Gamma_{h \to X X^\dag}}{\Gamma_h^{\rm tot}} 
\eeq
where $\Gamma_h^{\rm tot}$ is the total decay rate of SM Higgs.
Recent results from ATLAS and CMS collaborations are \cite{ATLAS,CMS}
\bea
\mu_{\rm ATLAS} &=& 1.43 \pm 0.21 \quad {\rm for} \ m_h = 125.5 \GeV \ ,
\\
\mu_{\rm CMS} &=& 0.8 \pm 0.14 \quad \hspace{0.5em}  {\rm for} \ m_h = 125.7 \GeV \ .
\eea
Hence the invisible decay of Higgs to dark matter can be consistent with CMS data 
only if  $\lambda_{HX} \ll 0.1$ or $m_X$ is very close to $m_h/2$.
On the other hand, if vacuum stability is imposed, such a small $\lambda_{HX}$ is excluded and only $m_X = \mathcal{O}(10^{2-3}) \GeV$ is allowed.
In this case, the production and decay rate of Higgs boson in our model are exactly 
the same as those of SM Higgs boson, since $H\rightarrow X X^\dagger$ is 
kinematically forbidden.  Therefore it is difficult to discriminate our model from SM in such a case.
In other words, if collider experiments shows any non-SM signature, our model is excluded.

It may be possible to search for a pair of dark matter production at the LHC or the ILC 
through $ e^+ e^- \rightarrow Z h^* \rightarrow Z ( X X^\dagger )$, or $WW$ fusion 
through 
\[
q \bar{q}  \rightarrow q \bar{q} h^* \rightarrow q \bar{q} X X^\dagger 
\]
with extra emissions of gluon or $\gamma$ from the initial or the final quark jets.
The detailed study of this channel will be beyond the scope of this paper, and will
be addressed elsewhere.

\section{Variations of the model}

Instead of our model analyzed in this paper, one can consider a simpler dark sector 
which contain either $X$ or $\psi$ only in addition to $\hat{B}'_\mu$.  
In these cases, renormalizable RH-neutrino portal interactions are not possible 
and leptogenesis from seesaw sector has nothing to do with the dark matter. 
If $\psi$ is absent and $X, X^\dagger$ are dark matters, the only change relative to 
our present model is that the current dark matter relic density should come from the 
thermal freeze-out of $X$-$X^\dag$ annihilation via  $\lambda_{HX}$ interaction.  
Hence the annihilation is fixed to be  \eq{sv-th} as usual.

If $X$ is absent and $\psi$ is the dark matter, one has to introduce a real SM-singlet 
scalar (say $S$) connecting the dark sector to the SM sector as the model discussed in 
Ref.~\cite{SFDM1}, so that the thermal freeze-out of $\psi$-$\bar{\psi}$ 
annihilation via the newly introduced interactions of $S$ provides a right amount of 
dark matter at present.  Otherwise $\psi$ and $\bar{\psi}$ would be overproduced 
due to the smallness of $\alpha_X$.  The physics of this model is nearly same as that   
discussed in Refs.~\cite{SFDM1,Baek:2012uj} modulo the effects of the dark 
interaction on structure formation and direct DM searches, as well as dark radiation 
from massless hidden photon.   The spin-independent cross section of the $\psi$ 
(or $\bar{\psi}$)-to-nucleon scattering via photon exchange is the same as the one 
in the case of $X$-$X^\dag$ dark matter, so the constraint on the kinetic mixing 
($\epsilon$) shown in the left panel  of Fig.~\ref{fig:lHX-Xenon-bound} is equally 
applicable to this case.  Higgs inflation is still possible in these variations as discussed  
in Ref.~\cite{Lebedev:2012zw}, since there are extra scalar fields $X$ or $S$ in either case.

Finally one could consider the case the $U(1)_X$ dark symmetry is spontaneously 
broken  by nonzero $\langle \phi \rangle \neq 0$. Then in this case there is a singlet
scalar from $\phi$ after $U(1)_X$ breaking, which will mix with the SM Higgs boson. 
Therefore there are two Higgs-like neutral scalar bosons after all, and both of
them have signal strengths universally suppressed from the SM value ``1''.  
If $\psi$ is the CDM, one needs a singlet scalar $S$ as a messenger, and this will
mix with the SM Higgs boson (and the remnant from $\phi$).  

In Table~2, we summarize the dark field contents, messengers, the particle identity 
of the dark matter (DM), the amount of dark radiation (DR) and the signal strengths 
of Higgs-like neutral scalar bosons (including the number of them) in various scenarios.
In all cases, there are additional scalar bosons (either $X$ or $\phi$ or both) which 
make Higgs inflation still viable for $m_H = 125$ GeV.   And the Higgs signal strength 
is smaller than ``1'' except for the scalar is the CDM with unbroken $U(1)_X$ dark 
symmetry. Especially $\mu_{i=1,2,(3)} < 1$ for fermion CDM, whether $U(1)_X$ is 
broken or not. Our conclusions on the Higgs signal strength are based on the 
assumption that there is only one Higgs doublet in the model.  If we include 
additional Higgs doublets or triplet Higgs, the Higgs portal would have richer structure,
and the signal strength will change completely and will vary depending on the Higgs 
decay channels. Also it should be possible to have a signal strength for 
$H\rightarrow \gamma\gamma$ channel greater than  ``1''  without difficulty.  


\begin{table}[htdp]
\begin{center}
\begin{tabular}{|c|c|c|c|c|c|}
\hline
Dark sector fields & $U(1)_X$ & Messenger & DM & Extra DR & $\mu_i$  
\\   \hline  
$\hat{B}'_\mu , X , \psi$ & Unbroken & $H^\dagger H  , \hat{B}'_{\mu\nu} \hat{B}^{\mu\nu} , N_R$ 
& $X$ & $\sim 0.08$ & $1~(i=1)$
\\
$\hat{B}'_\mu , X$ & Unbroken & $H^\dagger H , \hat{B}'_{\mu\nu} \hat{B}^{\mu\nu}$ 
& $X$ & $\sim 0.08$ & $1 ~(i=1)$
\\
$\hat{B}'_\mu , \psi$ & Unbroken & $H^\dagger H  , \hat{B}'_{\mu\nu} \hat{B}^{\mu\nu} ,  S$ 
& $\psi_X$ & $\sim 0.08$ & $< 1~(i=1,2)$
\\
\hline 
$\hat{B}'_\mu , X , \psi , \phi$ & Broken & 
$H^\dagger H  , \hat{B}'_{\mu\nu} \hat{B}^{\mu\nu} , N_R$ 
& $X$ or $\psi$ & $\sim 0$ & $< 1~(i=1,2)$
\\
$\hat{B}'_\mu , X , \phi$ & Broken 
& $H^\dagger H , \hat{B}'_{\mu\nu} \hat{B}^{\mu\nu}$ & $X$ & $\sim 0$ & $< 1 ~(i=1,2)$
\\
$\hat{B}'_\mu , \psi$ & Broken 
& $H^\dagger H  , \hat{B}'_{\mu\nu} \hat{B}^{\mu\nu} ,  S$ 
& $\psi$ & $\sim 0$ & $~~< 1~(i=1,2,3)$
\\   \hline
\end{tabular}
\end{center}
\caption{Dark fields in the hidden sector, messengers, dark matter (DM), the 
amount of dark radiation (DR), and the signal strength(s) of the $i$ scalar boson(s) 
($\mu_i$)  for unbroken or spontaneously broken (by $\langle \phi \rangle \neq 0$) 
$U(1)_X$ models considered 
in this work.  The number of Higgs-like neutral scalar bosons could be 1,2 or 3, 
depending on the  scenarios. }
\label{default}
\end{table}%

\section{Discussions of some miscellaneous issues}
\subsection{Comparison with other models}
Leptogenesis in our model is very similar to one in 
Ref.~\cite{Falkowski:2011xh} in that the RH neutrino decay is the origin of 
BAU and CDM observed today.
However there are a few important different aspects of our model compared with 
Ref.~\cite{Falkowski:2011xh}:
\begin{itemize}
\item Our lagrangian is based on local gauge symmetry $G_X = U(1)_X$ that 
guarantees that CDM is absolutely stable.   Assuming all the SM singlet fields are 
portals to the hidden sector DM, we are naturally led to the present model without 
any other {\it ad hoc} assumptions.
\item If stable, either $X$ or $\psi$ could be dark matter, but the smallness of the 
dark interaction that is required from observations of large scale structure does not 
allow the fermion dark matter $\psi$ in our model.
Hence $\psi$ should be able to decay.
The interaction to SM via RHN allows the decay of $\psi$ if $X$ is lighter than $\psi$.
Due to this process, dark matter is composed of $X$ and $X^\dag$, and becomes symmetric eventually irrespective of its origin (symmetric thermal or asymmetric non-thermal).
\item Because of the smallness of dark interaction, the freeze-out abundance of 
$\psi$ before decay is quite large and becomes main contribution to dark matter abundance of
$X$-$X^\dag$.   In other words, asymmetric production of dark matter does not play 
any significant role in our scenario, and the eventual relic density of dark matter is  
determined by thermal or non-thermal freeze-out of $X$-$X^\dag$ pair annihilation  
through Higgs portal $\lambda_{XH}$ terms. 
\item The decays of $\psi$ and $\bar{\psi}$ via RH neutrino portal contribute to the visible sector lepton number asymmetry even if there is no asymmetry between $\psi$ and $\bar{\psi}$.
The contribution can be the origin of the present baryon number asymmetry if the coupling of the dark $U(1)_X$ is small ($\alpha_X \lesssim 10^{-5}$) and $\psi$ is heavy enough ($m_\psi \gtrsim 10^3 \TeV$). 
\item Higgs inflation can be realized thanks to the existence of the portal interaction 
which is necessary for efficient pair annihilation of $X$ and $X^\dagger$ 
and for vacuum stability.  The large enough reheating temperature after inflation can 
set a proper initial condition for leptogenesis to work.
\item Since the dark symmetry is an unbroken gauge symmetry in our model, 
there is always dark radiation from massless hidden photon. 
This conclusion can be evaded, if the dark symmetry is an unbroken but confining symmetry like color gauge symmetry in  QCD          \cite{ko_hidden_qcd1,ko_hidden_qcd2,ko_hidden_qcd_proceeding1,
ko_hidden_qcd_proceeding2}.
In that case, CDM would be a composite hadron made of hidden sector quarks 
(similar to baryons or mesons in ordinary QCD without electroweak interacrtion) 
and would be absolutely stable.
\item The dark matter self-interaction caused by the massless dark photon can explain the small scale structure problem appearing in usual collisionless CDM scenario.
\end{itemize}

\subsection{Effects of nonrenormalizable operators}

Since our model has no Landau pole or vacuum instability up to Planck scale 
($M_{\rm Pl}$),  this model could be an ultimate theory up to Planck scale 
when we ignore the fine tuning problems related with the Higgs mass$^2$ or 
cosmological constant. 
Still there may be higher dimensional nonrenormalizable operators suppressed by  
some positive powers of $1/M_\pl$ originating from quantum gravity effects.  
Since the dark symmetry $U(1)_X$ is not broken, dark matter would be absolutely 
stable even in the presence of these higher dimensional operators 
\footnote{In case local dark symmetry is spontaneously broken, dark matter candidates 
may decay via nonrenormalizable operators. This will be discussed in detail in a
separate publication~\cite{progress_broken}.}.

In this section, let us list some dim-5 or dim-6 operators that are suppressed by one 
or two powers of  $1/M_\pl$ and contain either the RH neutrino or the dark  
fields \footnote{One can refer to Ref.~\cite{Buchmuller:1985jz,Grzadkowski:2010es} for dim-5 and dim-6 operators in the 
SM case.}, and discuss their effects on the results obtained in the previous sections.  
We still impose the local gauge invariance of the nonrenormalizable operators 
under local gauge symmetry transformations, 
$SU(3)_C \times SU(2)_L \times U(1)_Y \times U(1)_X$.

Dim-5 and dim-6  operators with $\overline{\psi} \psi$ will contribute to 
thermalization of $\psi$ and $\bar{\psi}$: 
\begin{eqnarray} 
{\rm dim-5}: & & \frac{1}{M_\pl} \overline{\psi} \psi H^\dagger H \ , \ \  
\frac{1}{M_\pl} \overline{\psi} \psi X^\dagger X \ , \ \  
\frac{1}{M_\pl}\overline{\psi} \sigma^{\mu\nu} \psi B_{\mu\nu} \ , \ \ 
\frac{1}{M_\pl}\overline{\psi} \sigma^{\mu\nu} \psi B'_{\mu\nu} 
\\
{\rm dim-6}: & & \frac{1}{M_\pl^2} \overline{\psi} \gamma_\mu \psi 
\overline{f} \gamma^\mu f \ , \ \   
\frac{i}{M_\pl^2} \overline{\psi} \gamma_\mu \psi 
\left[ H^\dagger D^\mu H - ( D^\mu H^\dagger ) H \right] 
\ , \ \   etc.
\end{eqnarray} 
where $f$ is the SM chiral fermion field.
The first two dim-5 operators above contribute to  $\psi \overline{\psi} \rightarrow 
H H^\dagger ,  X X^\dagger$, whose cross section is  estimated as 
\[
\sigma \sim \frac{1}{ 4\pi M_{\rm Planck}^2} 
\] 
in the limit $m_\psi \gg m_H, m_X$ which is legitimate in our model.  
This is far less than the cross section into a pair of massless dark photon, Eq.~(3.4), 
derived in Sec.~3.1 even if the dark gauge coupling is very small.  
The dim-6 and dim-7 operators will be even smaller.

Dim-6 operators including $X^\dagger X$ will contribute to thermalization of  
$X$ and $X^\dagger$ \footnote{There are no gauge invariant dim-5 operators 
involving $X^\dagger X$. }:
\begin{eqnarray} 
{\rm dim-6}: & & \frac{i}{M_\pl^2} 
\left[  X^\dagger ( D_\mu X ) - (D_\mu X^\dagger ) X \right] 
\overline{f} \gamma^\mu f  \ , \ \ 
\frac{1}{M_\pl^2} X^\dagger X {\cal O}_{\rm SM}^{(4)} \ , 
\\
&&  \frac{1}{M_\pl^2} 
\left[  X^\dagger ( D_\mu X ) - (D_\mu X^\dagger ) X \right] 
\left[ H^\dagger D^\mu H - ( D^\mu H^\dagger ) H \right] 
 \ , \ \  etc.
 \end{eqnarray} 
where ${\cal O}_{\rm SM}^{(4)} $ represents 
the dim-4 gauge invariant SM operators appearing in the SM lagrangian. 
The cross section for annihilation of $X$ and $X^\dagger$ from dim-6 operators 
is estimated as 
\[
\sigma (X X^\dagger \rightarrow {\rm SM ~particles}) 
\sim \frac{m_X^2}{4\pi M_\pl^4}
\]
which is totally negligible compared with the annihilation through renormalizable 
Higgs portal interaction involving $\lambda_{HX}$.

We do not show $1/M_\pl^2$-suppressed dim-6 operators for $\psi$ decay 
into $X l_{Li} H$, since it is far subdominant than the dim-6 operators generated by the 
virtual RH neutrinos which we already studied in Sec.~5. 

\section{Conclusion}
In this paper, we showed that, if the dark matter stability is guaranteed by an unbroken 
local dark symmetry with nonzero dark charge, renormalizable portal interactions of 
the RH neutrinos and SM Higgs ($H^\dagger H$) fix the minimal field contents of 
dark sector (a scalar $X$ and fermion $\psi$ as well as massless dark photon 
$\hat{B}'_\mu$) and allow very rich physics without conflicting with various 
phenomenological, astrophysical and cosmological constraints coming from the 
existence of massless dark photon.

The unbroken local dark symmetry is very strongly constrained by small and large scale 
structure formation, requiring the dark fine structure constant to be 
\beq
\alpha_X \lesssim 10^{-5} - 10^{-4}
\eeq
for $\mathcal{O}(10^{2-3}) \GeV$ scale mass of dark matter.  
On the other hand, the dark interaction can be the solution to the core/cusp 
and ``too big to fail'' problems 
of small scale structure in collisionless CDM scenarios.  The smallness of dark 
interaction could cause a danger of dark matter over-abundance. In our model, 
this potential danger is removed by RH neutrino portal which allows $\psi$ to decay 
to $X$, and Higgs portal which allows efficient dilution of the stable dark matter $X$ 
to get a right amount of dark matter relic density.   All these nice features are 
consequences of local dark gauge symmetry, and the assumption that all the SM 
singlet operators being portals to the dark sector.  The RH neutrino portal also allows 
production of dark sector asymmetry as leptogenesis in type-I seesaw model does in 
the visible sector, but the dark sector asymmetry eventually disappears as $\psi$ 
decays and does not play any significant role.
However, one should note that in our scenario eventual relic density of dark matter 
can be determined by thermal or non-thermal freeze-out, depending on the 
temperature when $\psi$ decays to $X$.  This allows wider range of dark matter 
annihilation cross section. 
Additionally, depending on $\alpha_X$ and the mass of $\psi$, the decays $\psi$ and $\bar{\psi}$ can be the origin of the present baryon number asymmetry irrespective of the possible asymmetry between $\psi$ and $\bar{\psi}$. 
Fig. 4 is a sketch of the brief thermal history of our leptogenesis, including the production of dark matter relics.

The Higgs portal interaction to dark scalar $X$ cures instability problem of the SM Higgs 
potential by loop effect.  So, by introducing large non-minimal gravitational couplings 
to scalar fields, it becomes possible to realize Higgs inflation whose high enough 
reheating temperature sets the initial condition for  leptogenesis in our model.
The portal interactions also make the dark sector be accessible by direct and/or 
indirect searches, which are consistent with the current bounds from various 
terrestrial experiments and observations in the sky.  

It turned out that the contribution of dark photon to the radiation density at present 
is about 8\% of the energy density of a massless neutrino. 
It is rather small, and still consistent with the present observation within 
$2$-$\sigma$ error. 
The smallness is originated from the fact that dark photon couples only to the dark sector fields and dark matter is decoupled from SM particles before QCD-phase transition.
 
Our model is a sort of the minimal model which has inflation, leptogenesis, absolutely 
stable dark matter and dark radiation, as well as seesaw mechanism for neutrino 
masses and mixings, modulo some variations described at section~8.
The basic principles for the model building were the local gauge symmetry working on 
the dark sector too, and the assumption of the SM singlet operators being portals 
to the dark sector. From these simple principles,  we could derive very rich physics 
results that are fully consistent with all the observations made so far.  
It is interesting to note that the Higgs property measurements will strongly constrain 
our model, since we predict that  the Higgs signal strength should be equal to or 
less than ``1" for all the decay channels of Higgs  boson. 


Our model could be extended in various directions. First of all, one can consider 
the case where the dark gauge symmetry is spontaneously broken. 
Depending on the local dark symmetry being broken fully or partially in case of 
non-Abelian dark gauge symmetry, one would have interesting phenomenological 
consequences in both dark matter and Higgs sectors.  
Secondly, it would be straightforward to extend our model to supersymmetric one. 
The qualitative features of SUSY models will be similar to this letter.
However there are two types of CDM. One in the MSSM sector, and the other in the 
other sector.  Therefore one could consider the $R$-parity in the MSSM sector is 
broken (either lepton number or baryon number).  Then the $G_X$-charged hidden 
sector dark matter will make the one in the universe.  These issues will be addressed 
in the future publication.

\begin{figure}[h] 
\centering
\includegraphics[width=16cm]{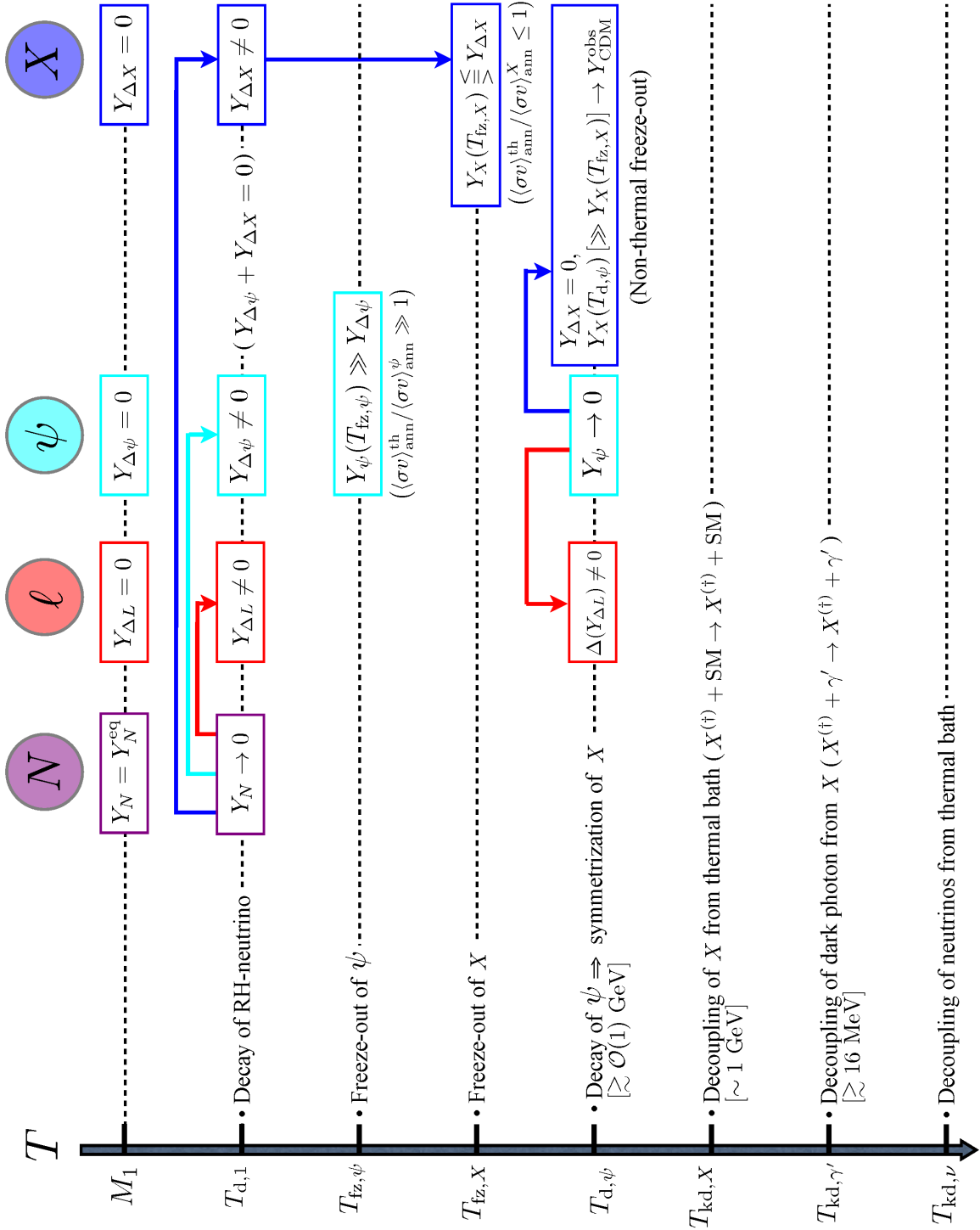}
\caption{Thermal history of the universe in our model}
\label{fig:history}
\end{figure}

\section*{Note Added}
After we submit this paper on the archive, we received a new result by 
the Planck Collaboration on the effective number of neutrino
species to be $N_{\rm eff} = 3.30 \pm 0.27$ at 68\% CL \cite{planck2013}, which is 
significantly lower than other previous results obtained by WMAP-9, 
SPT and ACT.  It is amusing to notice that the new Planck data is in 
perfect agreement with our prediction $(N_{\rm eff} = 3.130)$ in the model 
with unbroken $U(1)_X$ as well as other cases summarized in Table~2 in Sec.~8.

\section*{Acknowledgement}
We are grateful to Takeo Inami, Kenji Kadota, Hyunmin Lee and J.C. Park 
for useful discussions and comments. 
This work is supported in part by NRF Research Grant 
2012R1A2A1A01006053 (PK and SB), and by SRC program of NRF 
Grant No. 2009-0083526 funded by the Korea government(MSIP) (PK).

\appendix
\section{Thermally averaged cross sections}
In this Appendix, we collect the thermally averaged cross sections of dark matter pair annihilations.

\bea
\langle \sigma v \rangle_{X X^\dag \to \bar{f}f}
&=&
\frac{1}{32 \pi} N_c^f \lambda_{HX}^2 \frac{m_f^2}{\l( s - m_h^2 \r)^2 + m_h^2 \Gamma_h^2} \l( 1 - \frac{4 m_f^2}{s} \r)^{3/2} ,
\\
\langle \sigma v \rangle_{X X^\dag \to VV}
&=&
\frac{1}{64 \pi} \frac{\lambda_{HX}^2}{S} \frac{s}{\l( s - m_h^2 \r)^2 + m_h^2 \Gamma_h^2} \l[ 1 - 4 \frac{m_V^2}{s} + 12 \l( \frac{m_V^2}{s} \r)^2 \r] \l( 1 - \frac{4 m_V^2}{s} \r)^{1/2} ,
\\
\langle \sigma v \rangle_{X X^\dag \to hh}
&=&
\frac{1}{64 \pi s} \l( 1 - \frac{4 m_h^2}{s} \r)^{1/2} \int_{-1}^{1} d \cos \theta |A|^2
\eea 
where the symmetry factor is $S=(1,2)$ for $V=(W,Z)$ respectively, and 
\beq
|A|^2 = \frac{1}{4} \lambda_{HX}^2 \l| 1 - \frac{3 m_h^2}{\l( s - m_h^2 \r) + i m_h \Gamma_h} + \frac{1}{2} \frac{\lambda_{HX} v^2}{m_X^2 - t} + \frac{1}{2} \frac{\lambda_{HX} v^2}{m_X^2 - u} \r|^2 .
\eeq


\end{document}